\documentclass{appolb}
\usepackage{epsfig}
\begin{document}
\title{Tetraquark bound states in a constituent quark model and the
nature of the $a_0(980)$ and $f_0(980)$}
\author{J. Vijande $^{(1)}$, F. Fern\'andez $^{(1)}$, A. Valcarce $^{(1)}$ and
B. Silvestre-Brac $^{(2)}$
\address{$(1)$ Grupo de F\' \i sica Nuclear \\
Universidad de Salamanca, E-37008 Salamanca, Spain\\
$(2)$ Institut des Sciences Nucl\'eaires \\
53 Av. des Martyrs, F-38026 Grenoble cedex, France}
}
\maketitle
  
The pseudoscalar and vector meson ground states 
and their low-energy excited states 
have been understood as $q \bar q$ pairs. However, the
structure of the scalar mesons, ($J^{PC}=0^{++}$), remains
controversial nowadays. In the naive quark model, to construct
a positive parity state requires a unit of angular momentum
in a $q \bar q$ pair. Apparently, this takes an energy around 1 GeV since similar
meson states ($1^{++}$ and $2^{++}$) lie above 1.2 GeV.  
However a more complicated structure, like $q^2 \bar q^2$, suggested
twenty years ago by Jaffe \cite{jaf} can couple to $0^{++}$ without orbital excitation
and therefore could be a serious candidate to explain the structure of the lightest scalar mesons.

In this work we study tetraquark bound states
in the framework of the constituent quark model of Ref. \cite{nn1}, which has
been used for the description of non-strange two- and three-baryon systems and later
on applied to the hadron spectra. 
The model is based on the idea that between the scale of 
chiral symmetry breaking and the confinement scale, 
QCD may be formulated for the light quark sector 
as an effective theory of constituent quarks 
interacting through gluon and Goldstone
boson exchanges. For the heavy sector chiral symmetry 
is explicitly broken through the current 
quark masses and, as a consequence, the interaction reduces 
to confinement and gluon terms. 
Expressions of the interaction can be found elsewhere \cite{nn1}.

We will focus in two particular configurations. 
The first one will be the light-heavy 
states, $[(qq)(\bar Q \bar Q)]$, since they 
are the most prominent candidates to be bound 
under the strong interaction \cite{Pau,Sta}. 
The second one will be those tetraquarks
with the same quantum numbers as the scalar mesons \cite{acha}.
We solve the Schr\"odinger equation 
using a variational method where the spatial trial
wave function is a linear combination
of gaussians 

\begin{equation}
\label{eq1}
\Psi \propto e^{-a_i \vec x^{\,2}-b_i \vec y^{\,2}-c_i 
\vec z^{\,2}-d_i \vec x \cdot \vec y -e_i \vec x 
\cdot \vec z -f_i \vec y \cdot \vec z}
\end{equation}

\noindent 
$\vec x$, $\vec y$ and $\vec z$ being the tetraquark Jacobi coordinates 
and $a_i,b_i,c_i,d_i,e_i$ and $f_i$ the variational parameters.

The calculation is done using the full color-space
wave function, including the two possible color 
singlets, 
$\{\bar 3 3\}$ and $\{6 \bar 6\}$, 
despite of the fact that the $\{6 \bar 6\}$ 
component has been usually neglected \cite{Sta}.
The spatial exchange terms (those depending, for example, on
$\vec y \cdot \vec z$ in Eq. (\ref{eq1})) 
have also been usually neglected 
\cite{Sta}. If one makes such approximation
then, $[(qq)(\bar Q \bar Q)]$ 
$(I,S)=(0,0)$ and $(I,S)=(0,2)$ states  
would be forbidden \cite{ber2}.

We checked the importance of the spatial exchange
terms by comparing 
the results obtained with and 
without them. As a general trend we found
a contribution to the binding energy 
lower than 0.5\%. This is why the spatial
exchange terms will be only used
when studying the light-heavy states 
with $(I,S)=(0,0)$ or $(I,S)=(0,2)$, and will
be neglected in the other cases.
Most part of the model parameters have been fixed when
studying the $NN$ interaction and the baryon spectrum.
The remaining ones are fixed 
by adjusting the meson spectrum \cite{isg}.

In Table 1 we present our results for the different
flavor contents of the light-heavy systems.
We show the total energy ($E_T$) and 
the energy difference with respect to the lowest 
possible threshold,
$\Delta E\,=\,E_T(qq\bar Q\bar Q)-E_1(q\bar Q)-E_2(q\bar Q)$. 
Our calculation predicts only five stable tetraquarks. In 
particular one can see how the higher binding energies
are found in the $(I,S)=(0,1)$ channel, as has been 
usually predicted \cite{Sta}

The relative importance of both color channels is 
illustrated in Table 2. Being small in all cases,
one can see how the contribution of the $\{6\bar 6 \}$
color channel decreases when increasing the heavy quark
mass. We also observe a dependence on the spin of the
system. This behavior can be easily understood
taken into account that the one-gluon exchange 
is the responsible for the color channel mixing
and it depends on the inverse of the quark masses
and on the total spin of the system.

Concerning the possible tetraquark structure
of the scalar mesons we have studied systems 
with the same quantum numbers of the $a_0(980)$ and the $f_0(980)$. 
These mesons are $J^{PC}=0^{++}$ 
with strangeness 0, so they would correspond to a
tetraquark with $(I,S)=(0,0)$ for the $f_0$, and
$(I,S)=(1,0)$ for the $a_0$.
Experimental observations \cite{acha,peni} 
support the idea that the $f_0$ has a $s\bar s$ component, 
and therefore we will consider as first approach a
$[(qs)(\bar q \bar s)]$ structure
for the $f_0$ and $a_0$.

The obtained energies are the following: 
$E(I=0)=1282$ MeV and
$E(I=1)=1275$ MeV. We can see see how the energies 
are almost the same, as experimentally measured, but
the absolute value is too high for claiming that
these states correspond to the
experimental $a_0$ an $f_0$.

The origin of this discrepancy may be associated to our simplified ansatz for
the confining interaction. We have assumed that confinement has 
only two-body potentials, however in 
the tetraquark systems three- and four-body
terms are expected to play a role. To take into account 
these contributions, not
included in our model, we have modified the confinement parameter,
$a_c^{Tetraquark}=0.765*a_c^{Meson}$, obtaining $E(I=0)=987$ MeV and
$E(I=1)=981$ MeV, both nearly degenerated and with the right absolute value.

As a summary, the model studied  predicts a 
reasonable description of tetraquark
states in the light-heavy sector and 
supports the hypothesis of the $a_0$ and $f_0$ 
mesons having a tetraquark structure.

\begin{table}
\label{tab3}
\begin{tabular}{|ccc|ccccccc|}
\hline
&&&\multicolumn{7}{c|}{(I,S)}\\
&&&(0,0)&(1,0)&(0,1)&(1,1)&(0,2)&(1,2)&\\ 
\hline
&$[(qq)(\bar s\bar s)]$&$E_T$&2185&1660&1399&1810&2676&1894&\\
&&$\Delta E$&+1188&+663&+63&+474&+1002&+220&\\
&$[(qq)(\bar c\bar c)]$&$E_T$&4175&3947&3556&3975&4745&4001&\\
&&$\Delta E$&+438&+210&$-$254&+165&+862&+118&\\
&$[(qq)(\bar b\bar b)]$&$E_T$&10677&10493&10072&10498&11251&10509&\\
&&$\Delta E$&+120&$-$64&$-$509&$-$83&+646&$-$96&\\
\hline
\end{tabular}
\caption{Total ($E_T$) and binding energy ($\Delta E$) in MeV for the
light-heavy sector.}
\end{table}

\begin{table}
\begin{center}
\begin{tabular}{|cc|cc|ccc|}
\hline
&&\multicolumn{2}{|c|}{(I,S)=(1,0)}&\multicolumn{3}{c|}{(I,S)=(0,1)}\\
&& $\{\bar3 3\}$&\{$\bar3 3\}$ and $\{6\bar6\}$& $\{\bar3 3\}$&$
\{\bar3 3\}$ and $\{6\bar6\}$&\\
\hline
&$[(qq)(\bar s\bar s)]$&1762&1660&1422&1399&\\
&$[(qq)(\bar c\bar c)]$&3962&3947&3558&3556&\\
&$[(qq)(\bar b\bar b)]$&10493&10493&10072&10072&\\
\hline
\end{tabular}
\caption{Contribution of the different color channels in MeV.}
\end{center}
\end{table}


\begin{thebibliography}{99}
\bibitem{jaf} R.J. Jaffe. Phys. Rev. D15 (1977) 267.
\bibitem{nn1} F. Fern\'andez {\it et al}. J. Phys. G19 (1993) 2013. 
\bibitem{Pau} B. Silvestre-Brac, C. Semay. Z. Phys. C57 (1993) 273.
\bibitem{Sta} S. Pepin {\it et al.} Phys. Lett. B393 (1997) 119.
\bibitem{acha} N.N. Achasov. Nucl. Phys. A675 (2000) 279.
\bibitem{ber2} B. Silvestre-Brac. Phys. Rev. D46 (1992) 2179.
\bibitem{isg} J. Weinstein, N. Isgur. Phys. Rev. D27 (1983) 588.
\bibitem{peni} F. De Fazio, M.R. Pennington. Phys. Lett. B521 (2001) 15.
\end{thebibliography}
\end{document}